\documentclass[graybox]{svmult}

\usepackage{type1cm}        
\usepackage{makeidx}         
\usepackage{graphicx}        
\usepackage{multicol}        
\usepackage[bottom]{footmisc}

\usepackage{hyperref}
\usepackage{bm}
\usepackage{enumerate}

\usepackage{newtxtext}       %
\usepackage{newtxmath}       

\makeindex             

\allowdisplaybreaks

\begin{document}

\title*{Radio Pulsars as a Laboratory for Strong-field Gravity Tests}
\author{Lijing Shao}
\institute{Lijing Shao 
\at Kavli Institute for Astronomy and Astrophysics, Peking University, Beijing
100871, China \\ 
National Astronomical Observatories, Chinese Academy of Sciences, Beijing
100012, China \\
\email{lshao@pku.edu.cn}}
%
%
\maketitle

\abstract{General relativity offers a classical description to gravitation and
spacetime, and is a cornerstone for modern physics. It has passed a number of
empirical tests with flying colours, mostly in the weak-gravity regimes, but
nowadays also in the strong-gravity regimes. Radio pulsars provide one of the
earliest extrasolar laboratories for gravity tests. They,  in possession of
strongly self-gravitating bodies, i.e.\ neutron stars, are playing a unique role
in the studies of strong-field gravity.  Radio timing of binary pulsars enables
very precise measurements of system parameters, and the pulsar timing technology
is extremely sensitive to various types of changes in the orbital dynamics. If
an alternative gravity theory causes modifications to binary orbital evolution
with respect to general relativity, the theory prediction can be confronted with
timing results. In this chapter, we review the basic concepts in using radio
pulsars for strong-field gravity tests, with the aid of some recent examples in
this regard, including tests of gravitational dipolar radiation, massive gravity
theories, and the strong equivalence principle. With more sensitive radio
telescopes coming online, pulsars are to provide even more dedicated tests of
strong  gravity in the near future.}

\section{Introduction}
\label{sec:intro}

Pulsars are rotating magnetized neutron stars. On the one hand, due to their large
moment of inertia ($I \sim 10^{38}\,{\rm kg\,m}^2$) and usually small external
torque, their rotation is extremely stable.  If a pulsar sweeps a radiating beam
in the direction of the Earth, a radio pulse could be recorded using large-area telescopes
for each rotation. As fundamentally known in physics, such a periodic
signal can be viewed as a {\it clock}.  Therefore, pulsars are famously
recognized as astrophysical clocks in astronomy. Even better, thanks to a sophisticated
technique called pulsar timing~\cite{Taylor:1992kea}, pulsar astronomers can
{\it accurately} record a number of  periodic pulse signals. These pulses' times
of arrival are compared with atomic clocks at the telescope sites. Some of these
observations can be carried out and last for decades.  From a large number of
times of arrival of these pulse signals, the physical properties of pulsar
systems are inferred to a great precision~\cite{Lorimer:2005misc}. For example,
a recent study with sixteen years of timing data of the Double
Pulsar,\footnote{Currently, PSR~J0737$-$3039A/B is the only discovered double
neutron star system whose two neutron stars were both detected as
pulsars~\cite{Burgay:2003jj, Lyne:2004cj, Shao:2021iax}, known as Pulsar A and
Pulsar B.} PSR~J0737$-$3039A/B, gives the rotational frequency of pulsar A in
the binary system~\cite{Kramer:2021jcw},
\begin{align}
    \nu = 44.05406864196281(17) \,{\rm Hz} \,.
\end{align}
It has sixteen significant digits, and the numbers in the parenthesis give the
uncertainty of the last-two digits. Such a precision rivals the precision of
atomic clocks on the Earth~\cite{Hobbs:2012apa}, and also it  possibly calls for
an extension of the usual use of floating numbers in computer numerics for
future precision pulsar timing experiments. Pulsars are truly {\it precision clocks}.  

On the other hand, neutron stars are the
densest objects known that are made of standard-model materials. For such a
compact object, gravity plays a vital role in shaping its internal structure and
affecting its external dynamics. As explicitly demonstrated by Damour and
Esposito-Far\`ese~\cite{Damour:1993hw}, if gravity is described by an
alternative theory to the general relativity---in their case, a class of
scalar-tensor gravity theories---nonperturbative phase-transition-like
behaviours might happen for neutron stars, resulting in large deviations from
general relativity in the strong field of neutron stars~\cite{Damour:1996ke,
Esposito-Farese:2004azw, Sennett:2017lcx}.  These large deviations will manifest
in the timing data of pulsars in some way (cf. Sec.~\ref{sec:dipole}), and they
could provide smoking-gun signals for gravity theories regarding the
strong-field properties. Combining the strong-field nature of neutron stars and
the precision measurements of times of arrival, radio pulsars are truly ideal to
test alternative theories of gravity~\cite{Wex:2014nva, Shao:2016ezh,
Wex:2020ald}, augmenting what have been done in the weak field of the Solar
System~\cite{Will:2014kxa}, and complementing what are recently being performed
with gravitational waves~\cite{LIGOScientific:2019fpa, LIGOScientific:2020tif,
LIGOScientific:2021sio} and black hole shadows~\cite{EventHorizonTelescope:2019dse,
EventHorizonTelescope:2022xnr, EventHorizonTelescope:2022xqj}.

Currently, more than three thousands of radio pulsars are
discovered\footnote{\url{https://www.atnf.csiro.au/people/pulsar/psrcat/}}~\cite{Manchester:2004bp}.
The most useful subset of pulsars in testing alternative gravity theories are
millisecond pulsars in {\it clean} binaries.\footnote{In one case, a pulsar in a triple system, 
PSR~J0337+1715, provides the best limit on the strong equivalence 
principle~\cite{Archibald:2018oxs, Voisin:2020lqi, Shao:2016ubu}.}
Their times of arrival at
telescopes are imprinted with information from the following sources: 
\begin{enumerate}[(i)]
	\item the Solar system dynamics which affect the motion of radio telescopes;
    \item the binary dynamics which are resulted from the mutual gravitational
    interaction between the two binary components; and
    \item the interstellar medium which affects the propagation of radio waves
    in a frequency-dependent way, in terms of dispersion, scattering, and so on.
\end{enumerate}
A formalism, which includes the above effects and connects the proper time of
the pulse signals in the pulsar frame to the observed coordinate time at the
telescopes, is called a {\it pulsar timing model}. One of the widely used timing
models for binary pulsars is the Damour-Deruelle timing
model~\cite{Damour:1986ads}. It is a phenomenological model that applies to a
large set of  alternative gravity theories which are possibly being the
underlying theory for the binary's orbital motion. 

\begin{figure}[t]
    \centering
	\includegraphics[width=7cm]{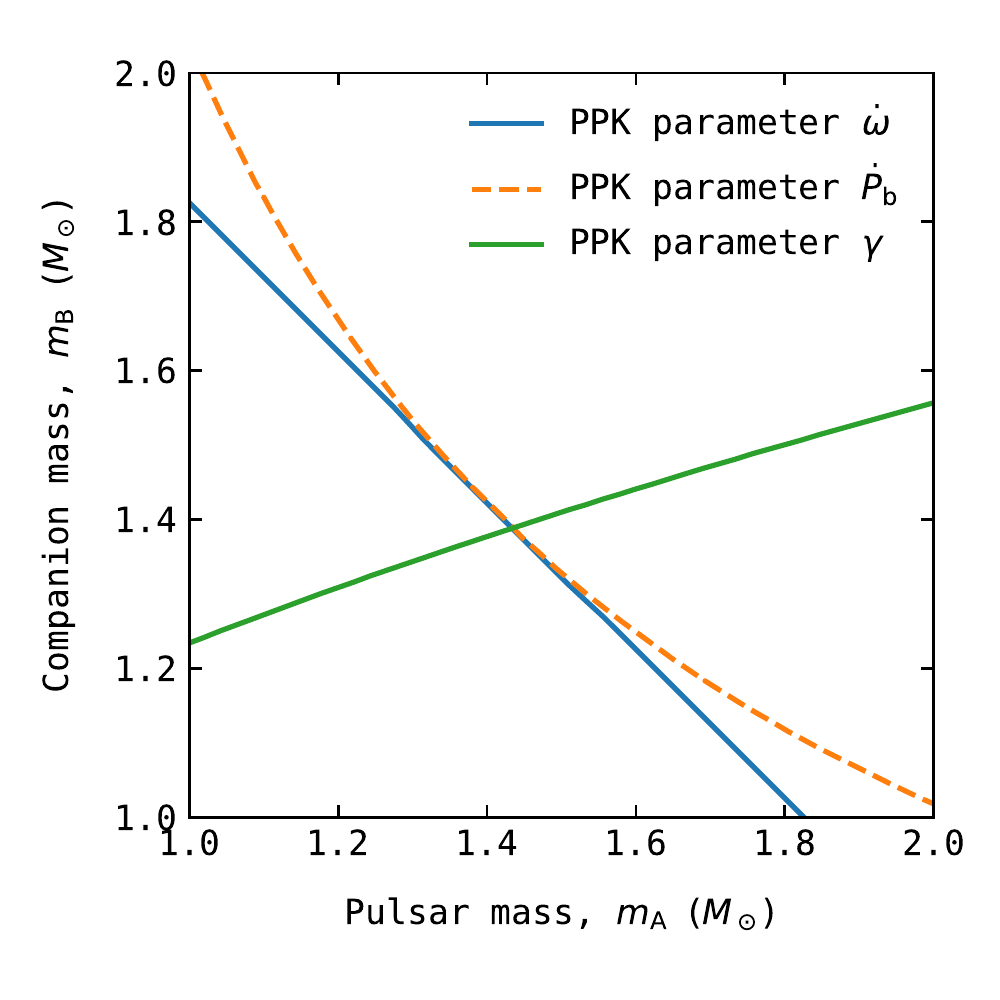}
	\caption{Consistency of general relativity in describing three measured PPK 
	parameters ($\dot\omega$, $\gamma$, and $\dot P_{\rm b}$) from PSR~B1913+16
	in the mass-mass diagram~\cite{Weisberg:2016jye}.  \label{fig:B1913}}
\end{figure}

In the Damour-Deruelle timing model, a handful of parameterized post-Keplerian
(PPK) parameters are introduced for generic Lorentz-invariant extensions of
gravity theories~\cite{Damour:1991rd}. The values of PPK parameters differ in
different gravity theories. Therefore, measurements of these PPK parameters can
be converted into constraints on parameters in the alternative gravity theories.
The most frequently used PPK parameters include $\dot\omega$, $\dot P_{\rm b}$,
$\gamma$, $r$, and $s$.  The PPK parameter $\dot\omega$ describes the periastron
advance of the binary orbit, the PPK parameter $\dot P_{\rm b}$ describes the
orbital period decay caused by the radiation of gravitational waves, the PPK
parameter $\gamma$ describes combined effects from the Doppler time delay and
gravitational time delay, and the  PPK parameters $(r,s)$ describe the Shapiro
time delay imprinted by the spacetime curvature of the companion star. The
values of these five PPK parameters in the general relativity are given
by~\cite{Damour:1986ads, Lorimer:2005misc},
\begin{align}
    \label{eq:omegadot}
    \dot{\omega} &=3\left(\frac{P_{\mathrm{b}}}{2 \pi}\right)^{-5 /
    3}\left(T_{\odot} M\right)^{2 / 3}\left(1-e^{2}\right)^{-1} \,, \\
    \label{eq:pbdot}
    \dot{P}_{\mathrm{b}} &=-\frac{192 \pi}{5}\left(\frac{P_{\mathrm{b}}}{2
    \pi}\right)^{-5 / 3}\left(1+\frac{73}{24} e^{2}+\frac{37}{96}
    e^{4}\right)\left(1-e^{2}\right)^{-7 / 2} T_{\odot}^{5 / 3} m_{\rm A} m_{\rm
    B} M^{-1 / 3} \,, \\ 
    \label{eq:gamma}
    \gamma &= e \left(\frac{P_{\mathrm{b}}}{2 \pi}\right)^{1 / 3} T_{\odot}^{2 /
    3}  M^{-4/3} m_{\rm B} (m_{\rm A} + 2 m_{\rm B})  \,, \\
    r &=T_{\odot} m_{\rm B} \,, \\
    \label{eq:shapiroS}
    s &=x\left(\frac{P_{\mathrm{b}}}{2 \pi}\right)^{-2 / 3} T_{\odot}^{-1 / 3}
    M^{2 / 3} m_{\rm B}^{-1} \,,
\end{align}
where $P_{\rm b}$ and $e$ are respectively the orbital period and orbital
eccentricity, $m_{\rm A}$ and $m_{\rm B}$ are the masses of the pulsar and its
companion in unit of the Solar mass ($M_\odot$), the total mass 
$M \equiv m_{\rm A} + m_{\rm B}$,
and $T_\odot \equiv G M_\odot / c^3 = 4.925490947 \mu$s.
Equations~(\ref{eq:omegadot})--(\ref{eq:shapiroS}) take different forms in
alternative gravity theories, often with dependence on the extra charges of the
binary components in the theory, e.g., these PPK parameters depend on scalar charges of the
pulsar and its companion in the scalar-tensor theory~\cite{Damour:1996ke}. In
pulsar-timing observation, each PPK parameter is {\it independently} measured.
Eventually, for a gravity theory to pass the tests from pulsar timing, it should
give consistent predictions to {\it all} the measured values of PPK parameters
with a {\it unique} set of physical parameters of the binary system. These
consistency checks are often illustrated in the mass-mass diagram.  For an
example, in Figure~\ref{fig:B1913} the measurements of three PPK parameters,
$\dot\omega$, $\gamma$, and $\dot P_{\rm b}$, from the Hulse-Taylor pulsar
PSR~B1913+16, give consistent component masses when the general relativistic
equations (\ref{eq:omegadot})--(\ref{eq:gamma}) are
used~\cite{Weisberg:2016jye}. Therefore, general relativity passes the tests
posed by the Hulse-Taylor pulsar~\cite{Weisberg:2016jye}.

In this following, we will give a few more concrete and recent examples where
binary pulsars play a key role in limiting alternative gravity theories,
including the gravitational dipolar radiation in the scalar-tensor gravity
(Sec.~\ref{sec:dipole}), two classes of massive gravity theories
(Sec.~\ref{sec:radiative}), and the strong equivalence principle
(Sec.~\ref{sec:conservative}). These examples are by no means complete, and
certainly reflect the somehow biased topics that the author is interested in. A
short perspective discussion is given in Sec.~\ref{sec:sum}. For more extensive
reviews on using radio pulsars for gravity tests, readers are referred to
Refs.~\cite{Stairs:2003eg, Wex:2014nva, Manchester:2015mda, Kramer:2016kwa,
Shao:2016ezh, Wex:2020ald}.

\section{Strong-field effects and gravitational dipolar radiation}
\label{sec:dipole}

Scalar-tensor gravity theories represent a well posed, healthy extension of Einstein's
general relativity by including a nonminimally coupled scalar field in the
Lagrangian of gravity~\cite{Brans:1961sx, Berti:2015itd, Will:2014kxa}. Shortly
after the first discovery of the Hulse-Taylor binary pulsar,
Eardley~\cite{Eardley:1975ads} pointed out that a gravitational dipolar
radiation could be used as a discriminant for such a class of gravity theories.
An extra dipolar radiation term can be tested with the PPK parameter $\dot P_b$.
Investigation along this line was boosted by the theoretical discovery that in a
slightly extended version of the original scalar-tensor gravity, nonperturbative
effects develop for certain neutron stars~\cite{Damour:1993hw, Damour:1996ke}.
The so-called {\it spontaneous scalarization} introduces a much enhanced
gravitational dipolar radiation for a scalarized neutron star in a binary. The
dipolar radiation in principle can even dominate over the quadrupolar radiation
predicted by the general relativity in binary pulsar observations [cf.
Eq.~(\ref{eq:pbdot})], but still keeping all weak-field gravity tests satisfied.
This enters the regime of {\it strong-field} gravity tests, where weak-field
tests have a rather limited power.

A general class of scalar-tensor gravity theories have the following action  in
the Einstein frame,
\begin{align}
    \label{eq:DEF}
    S=\frac{c^{4}}{16 \pi G_{*}} \int \frac{\mathrm{d}^{4} x}{c}
    \sqrt{-g_{*}}\left[R_{*}-2 g_{*}^{\mu \nu} \partial_{\mu} \varphi
    \partial_{\nu} \varphi-V(\varphi)\right]+S_{\rm m}\left[\psi_{\rm m} ;
    A^{2}(\varphi) g_{\mu \nu}^{*}\right] \,,
\end{align}
where  $g^{\mu\nu}_*$ and  $R_*$ are the metric tensor and  Ricci scalar
respectively, $\psi_{\rm m}$ collectively denotes standard-model matter fields,
$\varphi$ is an extra scalar field, and quantities with stars are in the
Einstein frame. The novel aspect lies in the fact that it is a conformal metric
$A^{2}(\varphi) g_{\mu \nu}^{*}$ instead of $g_{\mu \nu}^{*}$ itself that
couples to matter fields. Such a {\it nonminimal} coupling is important for the
discussions below.  

The class of scalar-tensor gravity theories carefully examined by 
Damour and Esposito-Far\`ese~\cite{Damour:1993hw, Damour:1996ke} has
\begin{align}
    V(\varphi) &=0 \,, \\
    A(\varphi) &=\exp \left(\beta_{0} \varphi^{2} / 2\right) \,, \\
    \alpha_{0} &=\beta_{0} \varphi_{0} \,, \label{eq:alpha0}
\end{align}
where $\varphi_0$ is the asymptotic value of $\varphi$ at infinity, and
$\alpha_0$ and $\beta_0$ are two theory parameters. This is the class of
scalar-tensor theories, sometimes denoted as $T_1(\alpha_0, \beta_0)$ and called
the {\it Damour-Esposito-Far\`ese theory}, that are most widely confronted with
pulsar observations~\cite{Freire:2012mg, Wex:2014nva, Shao:2017gwu,
Zhao:2022vig}.

\begin{figure}[t]
    \centering
	\includegraphics[width=8.5cm]{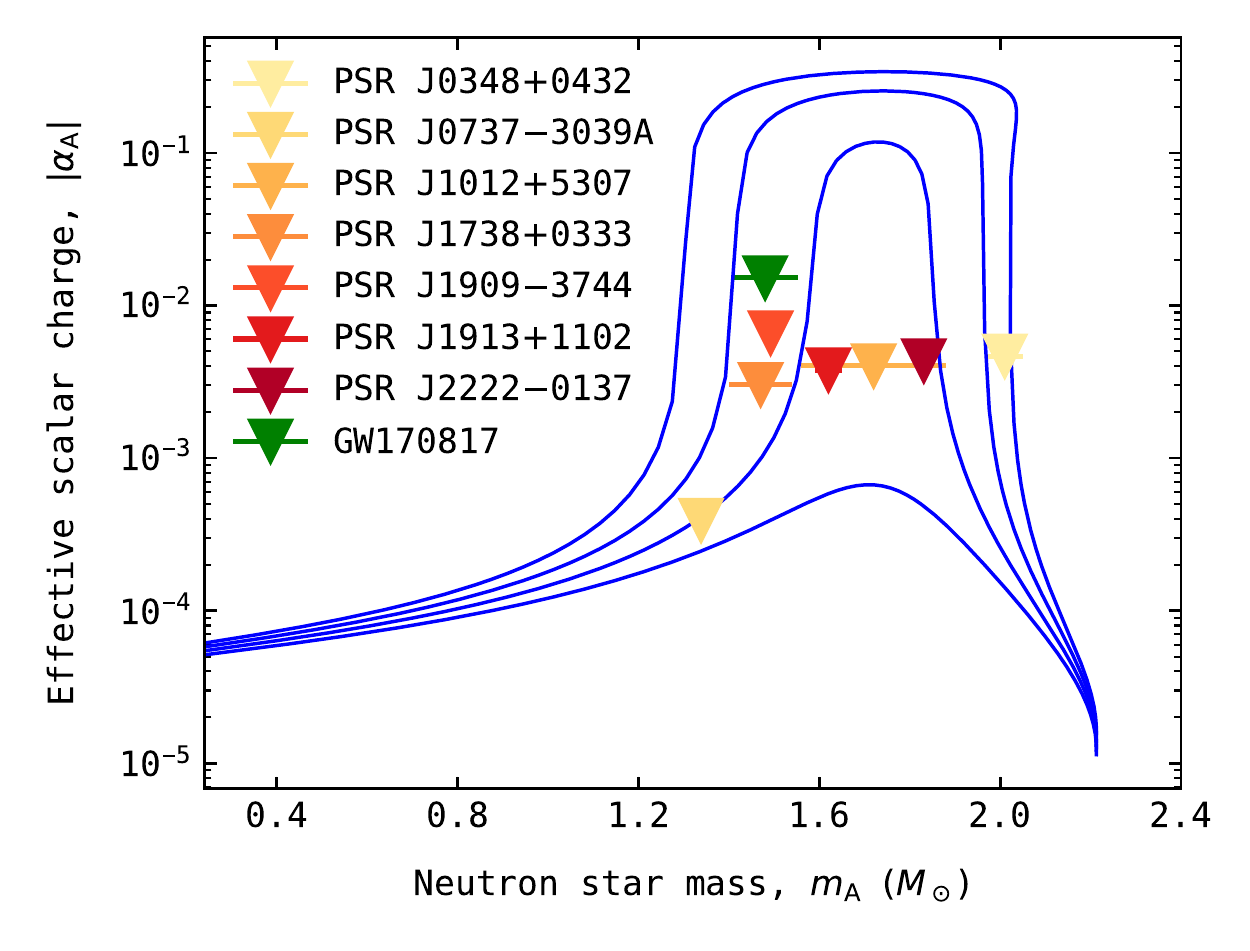}
	\caption{Blue curves show the effective scalar charge in the
	Damour-Esposito-Far\`ese scalar-tensor gravity theory with $|\alpha_0| =
	10^{-5}$ and, from top to bottom, $\beta_0 = -4.8, -4.6, -4.4, -4.2$. The
	{\tt AP4} equation of state is assumed in the calculation. Triangles show
	the observational bounds from binary pulsars~\cite{Shao:2017gwu,
	Zhao:2022vig} and gravitational waves~\cite{TheLIGOScientific:2017qsa,
	LIGOScientific:2018dkp} at the 90\% confidence level. The mass uncertainty
	for these neutron stars is indicated at the 68\% confidence level.
	\label{fig:scalarization}}
\end{figure}

By integrating the modified Tolman-Oppenheimer-Volkoff equations derived from
theory (\ref{eq:DEF}), one gets a boost in a neutron star's scalar charge when
its mass reaches a critical point. This phenomenon is understood from the
viewpoint of Landau's phase transition theory when a tachyonic instability kicks
in and a new branch of neutron star solutions with scalar charges are
energetically favored~\cite{Esposito-Farese:2004azw, Sennett:2017lcx, Khalil:2022sii}.  We
define the {\it effective scalar charge} of a neutron star~\cite{Damour:1993hw},
\begin{equation}
	\alpha_{\rm A} \equiv \frac{\partial \ln m_{\rm A}}{\partial \varphi_0} \,,
\end{equation}
which is a representative quantity characterizing the strength of deviation from
general relativity.  In Figure~\ref{fig:scalarization}, example curves for the
effective scalar charge as a function of neutron star mass are given in blue
lines   from top to bottom for $\beta_0 = -4.8, -4.6, -4.4, -4.2$, assuming the
{\tt AP4} equation of state and $|\alpha_0| = 10^{-5}$.  As we can easily seen,
indeed that for certain mass range of neutron stars, $|\alpha_{\rm A}|$ can be
very large while keeping its value very small in weak-gravity fields.

The emission of gravitational dipolar radiation in a binary pulsar is
proportional to the difference in the effective scalar couplings of the two
binary components A and B, and to the leading order, it contributes to an
additional decay rate of orbital period via~\cite{Damour:1996ke},
\begin{equation}
	\dot{P}_{\rm b}^{\text {dipole }}=-\frac{2 \pi G_{*}}{c^{3}}
	\left(1+\frac{e^{2}}{2}\right)\left(1-e^{2}\right)^{-5 / 2} \left(\frac{2
	\pi}{P_{\rm b}}\right) \frac{m_{\rm A} m_{\rm B}}{M}\left(\alpha_{\rm
	A}-\alpha_{\rm B}\right)^{2} \,.
\end{equation}
While neutron stars have significant scalar charges, white dwarfs, being
weak-field objects, are hardly different from their counterparts in general
relativity with a vanishingly small scalar charge $\alpha_{\rm B} \simeq
\alpha_0 \to 0$, where $\alpha_0$ is well constrained by Solar System weak-field
tests~\cite{Will:2014kxa}. Therefore, neutron-star white-dwarf binaries turn out
to be the most sensitive probe in this regard~\cite{Freire:2012mg,
Shao:2017gwu}. Recently, a new study~\cite{Zhao:2022vig} shows explicitly that
neutron-star neutron-star binaries with a significant difference in the masses
of binary components are also excellent laboratories. Therefore, to test the
gravitational dipolar radiation in scalar-tensor gravity, asymmetric binary
pulsars are needed\footnote{Unfortunately, we have not detected yet suitable
neutron-star black-hole binaries for this test, which are also potentially
very good testbeds~\cite{Liu:2014uka}.}~\cite{Wex:2014nva}. 

Some illustration for a specific equation of state, {\tt AP4}, is given in
Figure~\ref{fig:scalarization}, along with constraints on the gravitational
dipolar radiation from seven binary pulsars~\cite{Zhao:2022vig}: five
neutron-star white-dwarf binaries (PSRs~J0348+0432, J1012+5307, J1738+0333,
J1909$-$3744, and J2222$-$0137) and two asymmetric neutron-star neutron-star
binaries (PSRs~J0737$-$3039A and J1913+1102). For comparison, we also show a
constraint from the first binary neutron star merger observed via gravitational
waves~\cite{LIGOScientific:2018dkp}. In principle, the uncertainty in the
superanuclear neutron-star matter is entangled with strong-field gravity
tests~\cite{Shao:2019gjj}. Nevertheless, nowadays we have enough well-measured
binary pulsar systems to populate the whole mass range for neutron stars, and a
combined study~\cite{Shao:2017gwu, Zhao:2022vig} has verified that for each
reasonable equation of state, the possibility for spontaneous scalarization in
the Damour-Esposito-Far\`ese scalar-tensor gravity theory is very low.
Following the method developed by Shao et al.~\cite{Shao:2017gwu}, a dedicated Bayesian
parameter-estimation study combining the above-mentioned seven pulsar systems
has basically closed the possibility of developing spontaneous scalarization for
an effective scalar coupling larger than $10^{-2}$ for the theory given by
Eqs.~(\ref{eq:DEF})--(\ref{eq:alpha0}), no matter of the underlying
yet-uncertain equation of state for supranuclear neutron-star matters. 

It is worth to mention that, when performing Markov-chain Monte Carlo Bayesian
parameter estimation, the integration of the modified Tolman-Oppenheimer-Volkoff
equations needs to be carried out by more than millions of times on the fly thus
computationally expensive. Recently, reduced-order surrogate models, which
extract dominating features to represent accurate enough integration results, were
bulit to aid the speedup of the calculation~\cite{Zhao:2019suc, Guo:2021leu}.
The codes of these reduced-order surrogate models are publicly available at
\url{https://github.com/BenjaminDbb/pySTGROM} and
\url{https://github.com/mh-guo/pySTGROMX} for community use.

Although the original Damour-Esposito-Far\`ese scalar-tensor gravity theory is
disfavored by binary pulsar timing results, in further extended, generic
scalar-tensor gravity theories, neutron stars can still be scalarized. This is
particularly true for a massive scalar-tensor theory with $V(\varphi) \sim m^2
\varphi^2$ when the Compton wavelength of the scalar field is smaller than the
orbital separation of the binary~\cite{Ramazanoglu:2016kul, Yazadjiev:2016pcb,
Xu:2020vbs}. Basically the modification with respect to the general relativity
in the orbital dynamics is suppressed exponentially in a Yukawa fashion.
Fortuitously,  without giving much details, such kind of massive scalar-tensor
theories can be efficiently probed via the tidal deformability measurement in
gravitational waves~\cite{Hu:2020ubl, Hu:2021tyw, TheLIGOScientific:2017qsa}. In
this sense, a combination of pulsar timing data and gravitational wave data is
called for to probe a larger parameter space for scalar-tensor gravity
theories~\cite{Shao:2017gwu}.

In the past few years, other variants of scalar-tensor gravity theories triggered great
enthusiasm. Some of them not only give scalarized neutron stars, but also
scalarized black holes, in contrast to the {\it no-hair theorem}. A particularly 
interesting class of such theory includes
a topological Gauss-Bonnet term,
\begin{equation}
    \label{eq:gb}
	\mathcal{G}=R_{\alpha \beta \gamma \delta} R^{\alpha \beta
\gamma \delta}-4 R_{\alpha \beta} R^{\alpha \beta}+R^{2} \,,
\end{equation}
coupling to the scalar field~\cite{Doneva:2017bvd, Silva:2017uqg, Xu:2021kfh}.
In Eq.~(\ref{eq:gb}), $R_{\alpha \beta \gamma \delta}$ and $R_{\alpha \beta}$ are the Riemann
tensor and Ricci tensor respectively.  Preliminary constraints on the
scalar-Gauss-Bonnet gravity from binary pulsars are presented by Danchev et
al.~\cite{Danchev:2021tew}. This is a new field where observations of compact
objects including neutron stars and black holes are crucial to reveal the
strong-field information of gravitation.

\section{Radiative effects in massive gravity theories}
\label{sec:radiative}

Radiative tests from binary pulsars are powerful, as the related PPK parameter,
$\dot P_b$, can be very well measured from a long-term timing project on
suitable pulsars~\cite{Damour:1991rd}. This parameter improves with
observational time span $T_{\rm obs}$ quite fast, as $T^{-5/2}_{\rm obs}$. The
orbital decay rate $\dot P_b$ is not only useful for constraining the dipolar
gravitational wave emission, but also in other radiative aspects of gravitation,
for example, in constraining the extra radiation caused by the breaking down of
the Lorentz symmetry~\cite{Yagi:2013ava}, or a nonzero mass of
gravitons~\cite{Finn:2001qi, Miao:2019nhf, deRham:2012fw, Shao:2020fka}. Here we
give a brief introduction to the latter.

In general relativity, the hypothetical quantum particle for gravity, graviton,
is a massless spin-2 particle. However, massive gravity theories are found to provide
interesting phenomena related to the evolution of the Universe, e.g. the
accelerated expansion and dark energy~\cite{deRham:2012fw}. Therefore, probing
the upper bounds of the graviton mass is fundamentally important to field theories
and cosmology studies, and it is one of the central topics in gravitational
physics.

One of the early study of using binary pulsars to test the graviton mass was
performed by Finn and Sutton in 2002~\cite{Finn:2001qi}. They investigated a
linearized gravity with a massive graviton with the action,
\begin{align}
    \label{eq:finnsutton}
    S= \frac{1}{64 \pi} \int \mathrm{d}^{4} x\bigg[ & \partial_{\lambda} h_{\mu
    \nu} \partial^{\lambda} h^{\mu \nu}-2 \partial^{\nu} h_{\mu \nu}
    \partial_{\lambda} h^{\mu \lambda}+2 \partial^{\nu} h_{\mu \nu}
    \partial^{\mu} h \nonumber \\
    & -\partial^{\mu} h \partial_{\mu} h-32 \pi h_{\mu \nu} T^{\mu
    \nu}+m_{\rm g}^{2}\left(h_{\mu \nu} h^{\mu \nu}-\frac{1}{2} h^{2}\right)\bigg] \,,
\end{align}
where the last term gives a unique graviton mass under certain
conditions\footnote{The conditions are that (i) the wave equation takes a
standard form 
\begin{equation}
	\left(\square-m_{\rm g}^{2}\right) \bar{h}_{\mu \nu}+16 \pi T_{\mu \nu}=0 \,,
\end{equation}
and the theory recovers the general relativity in the limit when $m_{\rm g} \to
0$, namely, there is no van Dam-Veltman-Zakharov
discontinuity~\cite{Finn:2001qi}.}~\cite{Finn:2001qi} while the others are just linearized
expansions from the Einstein-Hilbert action with $h_{\mu\nu} \equiv g_{\mu\nu} -
\eta_{\mu\nu}$ and $h \equiv h^{\mu}_{~~\mu}$. It was shown that extra
gravitational wave radiation exists in theory~(\ref{eq:finnsutton}), which
results in a fractional change in the orbital decay rate, by~\cite{Finn:2001qi}
\begin{align}
    \frac{\dot{P}_{\rm b}-\dot{P}_{\rm b}^{\mathrm{GR}}}{\dot{P}_{\rm
    b}^{\mathrm{GR}}}=\frac{5}{24}
    \frac{\left(1-e^{2}\right)^{3}}{1+\frac{73}{24} e^{2}+\frac{37}{96}
    e^{4}}\left(\frac{P_{\rm b}}{2 \pi \hbar}\right)^{2} m_{\rm g}^{2} \,.
\end{align}
Here $\dot{P}_{\rm b}^{\mathrm{GR}}$ is the value predicted by the general
relativity in Eq.~(\ref{eq:pbdot}).  Notice that the fractional change is
proportional to $\propto P_{\rm b}^2 m_{\rm g}^2$. Therefore, if the precision of $\dot
P_{\rm b}$ is given, binary pulsars with larger orbits have a larger figure of
merit for the test. However, usually, the precision of $\dot P_{\rm b}$
crucially depends on the orbital size, and it turns out that, still, binary
pulsars with smaller orbits have a larger figure of merit.

The most recent constraint in this Finn-Sutton framework was provided by a
combination of multiple best-timed binary pulsars with a Bayesian statistical
treatment.  A collection of nine best-timed binary pulsars (PSRs~J0348+0432,
J0737$-$3039, J1012+5307, B1534+12, J1713+0747, J1738+0333, J1909$-$3744,
B1913+16, and J2222$-$0137) provide a tight bound on the graviton mass,
\begin{align}
    m_{g}<5.2 \times 10^{-21} \, \mathrm{eV} / c^{2}, \quad(90 \% \text { C.L.})
    \,,
\end{align}
using a uniform prior in $\ln m_{\rm g}$~\cite{Miao:2019nhf}.  This limit is not
the strongest limit on the graviton mass~\cite{deRham:2016nuf}. However, from a
theoretical point of view, it is a bound from binary orbital dynamics,
complementary to, e.g.\ the kinematic dispersion-relation tests from the
LIGO/Virgo/KAGRA observation of gravitational
waves~\cite{LIGOScientific:2021sio}. It is worth mentioning that the
theory~(\ref{eq:finnsutton}) has some drawbacks including ghosts and
instability~\cite{Finn:2001qi, deRham:2016nuf}, and here it is only used as a
strawman target for illustration.

It is interesting to note, that in different massive gravity theories, the
dependence of the extra radiation on the graviton mass is in general different.
It depends on the specifics of the illustrated gravity theory. This is due to
the deep fundamental principles in the designs of a number of variants of
massive gravity theories.  For example, in a cosmologically motivated massive
gravity theory, known as the {\it cubic Galileon theory} with the
action~\cite{deRham:2012fw},
\begin{align}
    \label{eq:CG}
    S=\int \mathrm{d}^{4} x\left[-\frac{1}{4} h^{\mu \nu}(\mathcal{E} h)_{\mu
    \nu}+\frac{h^{\mu \nu} T_{\mu \nu}}{2
    M_{\mathrm{Pl}}}-\frac{3}{4}\left(\partial \varphi
    \right)^{2}\left(1+\frac{1}{3 m_{\rm g}^2 M_{\mathrm{Pl}} } \square \varphi
    \right)+\frac{\varphi T}{2 M_{\mathrm{Pl}}}\right] \,,
\end{align}
the specific way of the addition of the scalar field $\varphi$ introduces the
so-called {\it screening mechanism}, thus avoids the stringent constraints from
the Solar System, yet provides important changes to the cosmological evolution.
In the action (\ref{eq:CG}), $\varphi$ is the Galileon scalar field, $T_{\mu
\nu}$ is the matter energy-momentum tensor, $T \equiv T^\mu_{~~\mu}$, 
$M_{\rm Pl}$ is the Planck mass, and
\begin{equation}
	(\mathcal{E}
h)_{\mu \nu} \equiv-\frac{1}{2} \square h_{\mu \nu}+\cdots
\end{equation}
is the Lichnerowicz operator. For a central
massive body with mass $M$, the {\it screening radius} is $r_\star = \big( M /
16 m_{\rm g}^2 M_{\rm Pl}^2 \big)^{1/3}$, within which, the theory exhibits
strong couplings and it reduces to the canonical gravity.

\begin{figure}[t]
    \centering
	\includegraphics[width=7cm]{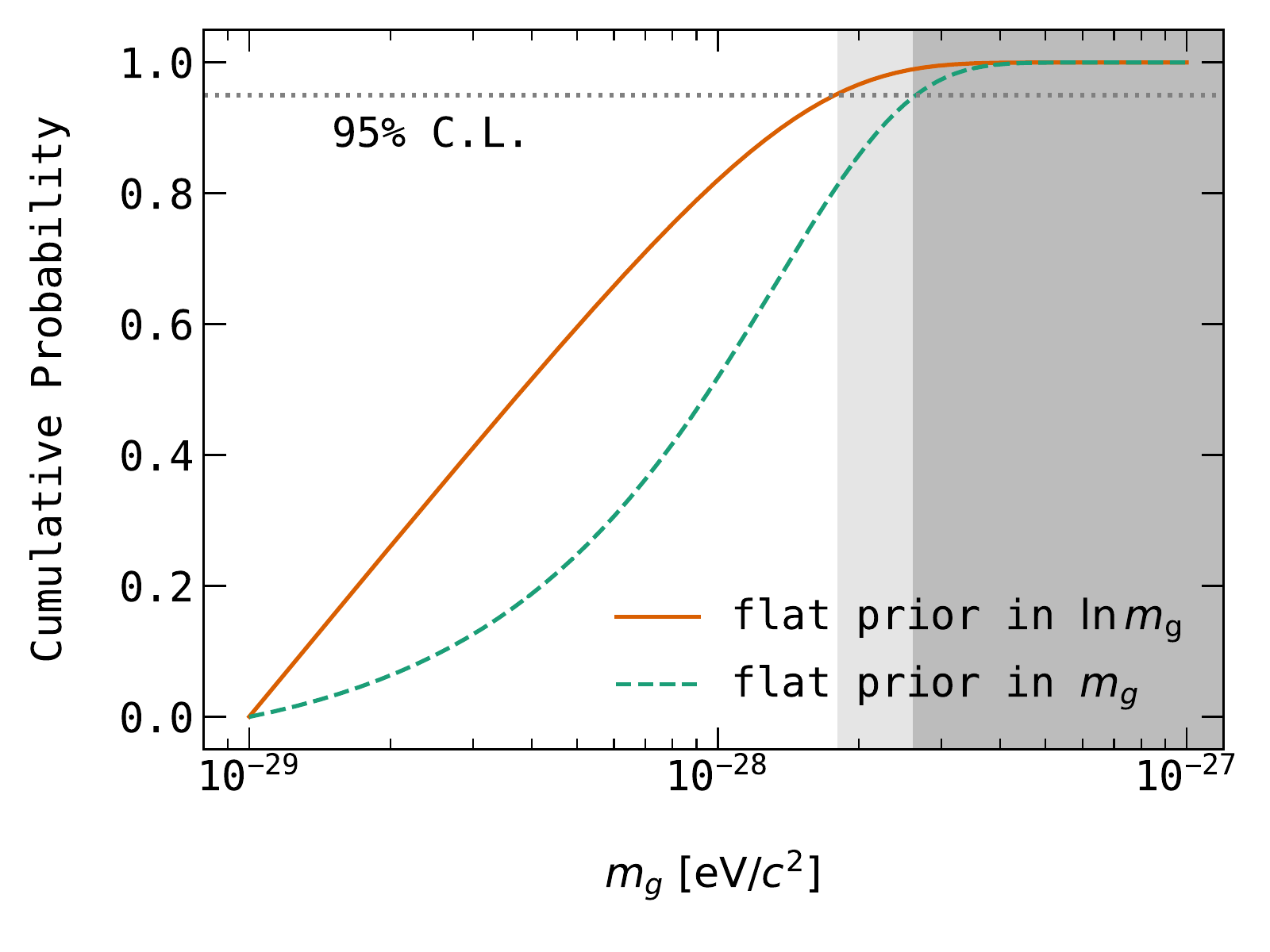}
	\caption{Cumulative probability for the graviton mass with two different
	priors in the cubic Galileon theory~\cite{Shao:2020fka}. Shaded regions show
	the excluded graviton mass values at the 95\% confidence level.
	\label{fig:galileon}}
\end{figure}

According to de Rham et al.~\cite{deRham:2012fw}, though with a screening
mechanism to suppress modification at the high density region within $r_\star$,
this cubic Galileon theory predicts a different scaling behaviour for the
gravitational radiation.  For a system with a typical length scale $L$, the {\it
fifth-force} suppression factor is $\sim \big( L/r_\star \big)^{3/2}$, and the
suppression factor for the gravitational radiation is $\sim (P_{\rm b} /
r_\star)^{3/2}$. As for a binary system, one has $L \sim v P_{\rm b}$ where $v$
is a characteristic velocity.  Therefore, the gravitational radiation is,
compared with the fifth force, {\it less} suppressed by a factor of $v^{3/2}$,
and it provides a valuable window to look for evidence of this theory via
radiative channels, for example, in binary pulsar systems.

Analytic radiative powers were worked out by de Rham et
al.~\cite{deRham:2012fw}, and the extra radiative channels include monopolar radiation, dipolar
radiation, and quadrupolar radiation. For binary pulsar systems with different
orbital periods and orbital eccentricities, the dominate radiation channel can
be different~\cite{Shao:2020fka}.  For the current set of binary pulsars, 
the quadrupole radiation is the dominating factor among the extra 
channels~\cite{deRham:2012fw, Shao:2020fka}.

The most up-to-date constraint from binary pulsars is
\begin{align}
    m_{g}<2 \times 10^{-28} \, \mathrm{eV} / c^{2}, \quad(95 \% \text { C.L.})
    \,, \label{eq:CG:limit}
\end{align}
for the cubic Galileon theory, and the cumulative probability distributions of the
graviton mass are given in Figure~\ref{fig:galileon} for two different priors~\cite{Shao:2020fka}. Such a
tight constraint was obtained from the combination of fourteen best-timed binary
pulsar systems, including PSRs~J0348+0432, J0437$-$4715, J0613$-$0200,
J0737$-$3039, J1012+5307, J1022+1001, J1141$-$6545, B1534+12, J1713+0747,
J1738+0333, J1756$-$2251, J1909$-$3744, B1913+16, and J2222$-$0137. One should
keep in mind that, the limit~(\ref{eq:CG:limit}) is theory specific, and in this
situation, only applies to the cubic Galileon theory given in Eq.~(\ref{eq:CG}).
Nonetheless, it provides an interesting example that for a gravity theory
designed for cosmological  purposes at corresponding lengthscales, binary pulsar 
systems with astronomical lengthscales still provide intriguing and useful bounds.
It is an illustration of using binary pulsars in the studies of cosmology by examining the 
modification to binary orbits brought by a cosmologically-motivated modified gravity.

\section{Strong equivalence principle and dark matters}
\label{sec:conservative}

Binary pulsars are not only useful for the radiative tests introduced in the
above sections, they also provide superb limiting power in the conservative
aspects of gravitational dynamics for orbital evolutions. Below we introduce an example of examining
the strong equivalence principle via the conservative dynamics of binary
pulsars~\cite{Damour:1991rq, Zhu:2018etc}, and its extension to test certain
interesting properties of dark matters~\cite{Wagner:2012ui, Shao:2018klg}.

\begin{figure}[t]
    \centering
	\includegraphics[width=6cm]{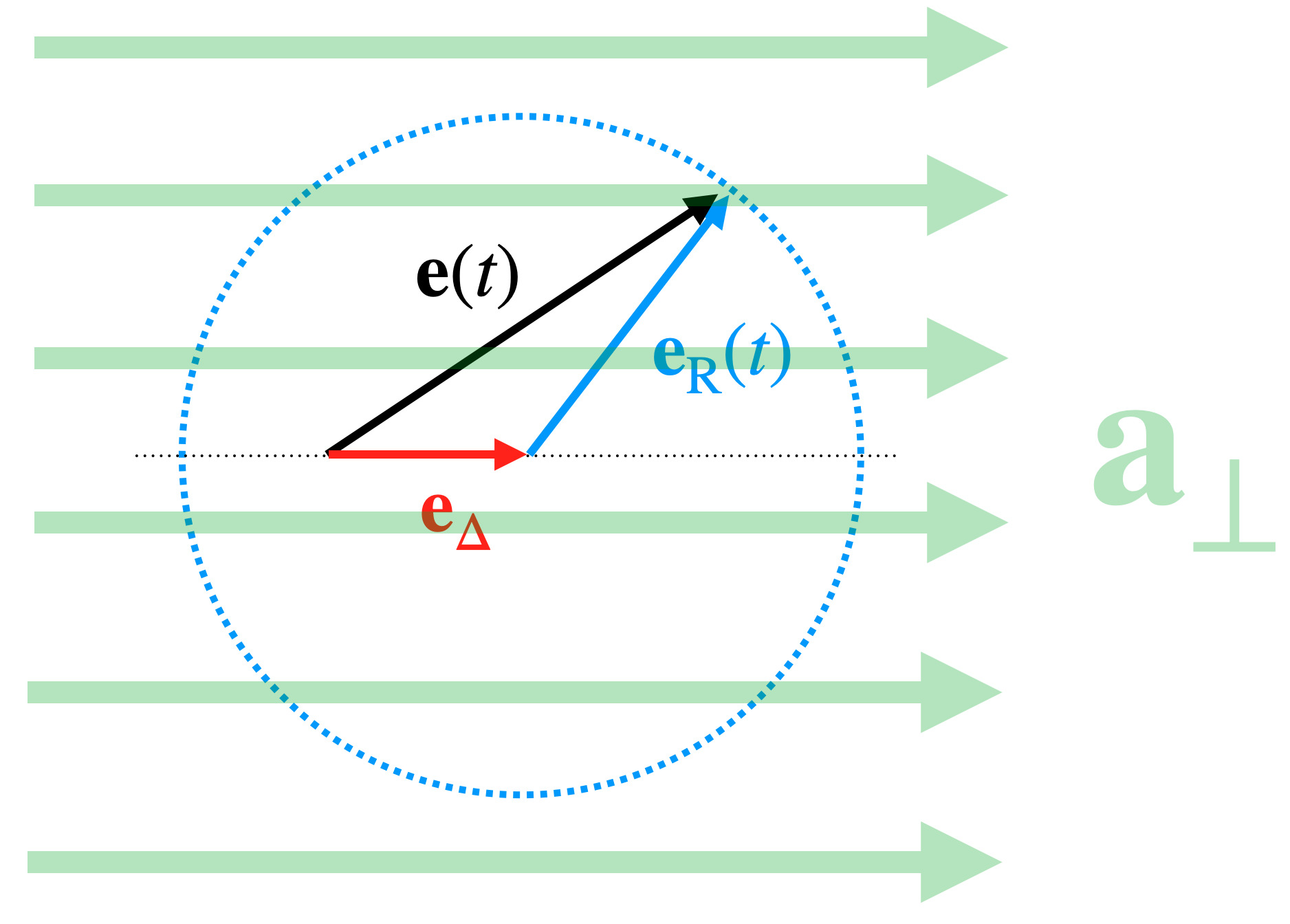}
	\caption{Graphical illustration of the time-varying orbital eccentricity
	vector, $\bm{e}(t)$,  for a binary pulsar, in the presence of strong
	equivalence principle violation~\cite{Damour:1991rq}. The orbital
	eccentricity vector evolves according to $\bm{e}(t) = \bm{e}_\Delta +
	\bm{e}_{\rm R}(t)$, where $\bm{e}_{\rm R}(t)$ is the usual precessing
	eccentricity vector in the general relativity, and the {\it constant}
	abnormal eccentricity is in the direction of $\bm{a}_\perp$, which is the projection
	of external Galactic acceleration in the orbital plane.  \label{fig:sep}}
\end{figure}

As discovered by Damour and Sch\"afer~\cite{Damour:1991rq}, a perturbed binary
orbit with an equivalence-principle-violating abnormal acceleration has a
characteristic evolution in its orbital elements. The notable change is the
appearance of a {\it vectorized superposition} of two eccentricity vectors for the real
orbital eccentricity.  It
provides a graphical understanding of the underlying dynamics for a binary in presence of
equivalence principle violations. The {\it real}
orbital eccentricity vector, $\bm{e}(t)$, is an addition of a rotating normal
eccentricity vector, $\bm{e}_{\rm R}(t)$, in its post-Newtonian fashion, and an
extra abnormal eccentricity vector, $\bm{e}_{\Delta}$, which is time independent
and whose length is proportional to the E\"otv\"os parameter, $\Delta$,
describing the violation of the equivalence principle. If $\Delta = 0$, the abnormal 
eccentricity vector $\bm{e}_{\Delta} = 0$ and it
returns to the precessing case in the general relavitity. A graphical illustration
is given in Figure~\ref{fig:sep}. As we discussed in Sec.~\ref{sec:intro}, the
pulsar timing technique is very sensitive to tiny changes in the orbit, and such
a change can be captured in pulsar timing data~\cite{Damour:1991rq}.

At the beginning, such a scenario was applied to a few binary pulsars in a statistical
sense by marginalizing  over some unknown angles to obtain constraints on 
the violation of the equivalence principle~\cite{Damour:1991rq}. Later it
was implemented to a handful of binary pulsars with an improved statistical
methodology to better account for the movements of binary pulsars in the 
Milky Way~\cite{Wex:2014nva}.  Then, with better data and more information
about binary pulsar systems, a direct method was developed~\cite{Freire:2012nb}.
The direct method not only can constrain the equivalence principle violation, 
but in principle can detect it if it exists.

The most stringent limit using binary pulsars comes from a precisely timed 
long-orbital-period binary
pulsar, PSR~J1713+0747~\cite{Zhu:2018etc}, as larger orbits have higher figures
of merit in such a test~\cite{Damour:1991rq}.  Using the improved direct method,
the limit on the E\"otv\"os parameter from PSR~J1713+0747 is~\cite{Zhu:2018etc},
\begin{align}
    |\Delta |<2 \times 10^{-3} \,, \quad(95 \% \text { C.L.})
    \,. \label{eq:Delta}
\end{align}
Though it is much less limiting than the constraint earlier obtained from the
Solar System~\cite{Wagner:2012ui,Will:2014kxa}, the limit~(\ref{eq:Delta})
encodes strong-field effects. For example, in the case of the aforementioned
scalar-tensor gravity, the strong-field version of E\"otv\"os parameter will be
very different from its weak-field counterpart~\cite{Freire:2012nb}.  Therefore,
such a limit from neutron stars is a {\it standalone} bound and applicable to the
strong version of equivalence principle~\cite{Wex:2014nva, Shao:2016ezh}.

The limit~(\ref{eq:Delta}) is not only interesting to gravitational physics, it
also has its value when we look at it from a different angle. As we now know,
the binary pulsar is actually immersed in the ocean of dark matters in the Milky
Way. As we have not really understood what the very nature of dark matter is, 
the above method for testing the equivalence principle
provides a non-traditional probe to dark matter's properties. Shao et
al.~\cite{Shao:2018klg} proposed a method where such a limit, with a proper
handle, can be converted to the interaction properties between dark matters and
ordinary matters. 

If there is a {\it long-range} {\it fifth force} between dark matter particles
and ordinary matter fields, as many field theories will
suggest~\cite{Wagner:2012ui}, it is likely to introduce an {\it apparent}
violation of the strong equivalence principle if we have not taken the fifth
force into account in our standard assumptions.  The role of the Galactic
acceleration in Figure~\ref{fig:sep}, whose projection on the orbital plane is
$\bm{a}_\perp$, is replaced by the attraction of dark matters to the binary
system. The difference in the acceleration to two binary components (a neutron
star and a white dwarf in the case of PSR~J1713+0747), described by $\Delta$, is
replaced by a quantity related to the long-range fifth-force between dark
matters and ordinary standard-model matters~\cite{Shao:2018klg}.

Detailed analysis of PSR~J1713+0747~\cite{Shao:2018klg} took into consideration
of the Galactic distribution of dark matters, and gave a very different bound in
nature that could be obtained from terrestrial experiments~\cite{Wagner:2012ui}.
The current observational data of PSR~J1713+0747 already imply that, if there is such a
long-range fifth force between dark matters and ordinary matters, its magnitude
should be no more than 1\% of the gravitational force between them. Such a limit
provides a useful complement to other types of dark-matter experiments, which
are usually looking for {\it short-range} forces between the hypothesized
dark-matter particles and the standard-model particles~\cite{Wagner:2012ui}, including the searches
in underground laboratories, particle colliders, and X-ray/$\gamma$-ray observations via
 high-energy satellites. 

\section{Summary}
\label{sec:sum}

In this chapter, we present some basic concepts of using binary pulsars as {\it
fundamental clocks} in a curved spacetime to probe various types of modifications
to the binary orbits. These modifications could have been caused by a modified
gravity theory or some other new physics like a long-range fifth force between dark
matters and ordinary matters. As pulsar timing provides us with very {\it
accurate} measurements, it puts constraints on tiny changes caused by an
alternative gravity theory other than the general relativity. Moreover, neutron
stars are intrinsically {\it strong-gravity} objects, and nonperturbative
aspects of the strong-field gravity can also be studied via radio pulsar experiments. Actually,
quite many strong-field limits are still best provided by pulsar timing
experiments, even nowadays in presence of new types of observations like
gravitational waves and black hole shadows. A careful study shows that the
limits from pulsar timing are actually complementary to those from gravitational
wave detections and black hole shadows~\cite{Shao:2017gwu, EventHorizonTelescope:2022xqj}.
Proper combinations of these strong-gravity experiments could provide a more complete landscape 
to gravitation in the strong-field .

Solely focusing on the radio pulsar side, the timing experiments can be carried out
for decades, in particular for some interesting systems like the Hulse-Taylor 
pulsar PSR~B1913+16~\cite{Weisberg:2016jye} and the Double Pulsar
PSR~J0737$-$3039A/B~\cite{Kramer:2021jcw}. Long-term observations improve the
precision of PPK parameters with the observational time span $T_{\rm obs}$. For
examples, the precision in the orbital decay parameter, $\dot P_{\rm b}$, improves very fast, as
$T_{\rm obs}^{-5/2}$, and the precision in the periastron advance rate, $\dot \omega$, improves
as $T_{\rm obs}^{-3/2}$. Furthermore, the sensitivity of radio telescopes is
also improving, notably with the Five-hundred-meter Aperture Spherical Telescope
in China~\cite{Jiang:2019rnj, Lu:2019gsr} and the Square Kilometre Array in
South Africa and Australia~\cite{Shao:2014wja, Bull:2018lat}. The former has
already been operating for a couple of years, while the latter has also entered
the construction phase recently. The improvement in the sensitivity of radio
telescopes directly converts to improvements in the timing precision. Therefore,
the real improvement for PPK parameters is faster than the theoretical power law
predictions. Last but not the least, radio telescopes are also continuously
discovering new pulsar systems, and some of these systems with suitable system
properties will contribute to strong-field gravity tests. We are even
looking forward to discovering yet-undetected binary pulsar systems like 
neutron-star black-hole binaries with short orbital periods $P_{\rm b} \lesssim 1$\,day
or pulsars around the Sgr\,A$^\ast$ black hole with orbital 
periods $P_{\rm b} \lesssim 10$\,years~\cite{Liu:2014uka,
Bower:2018mta, Bower:2019ads}, which will provide completely new gravity tests
in the strong-field regimes~\cite{Kramer:2004hd}.

\begin{acknowledgement}
We are grateful to the 740$^{\rm th}$ WE-Heraeus-Seminar ``Experimental Tests
and Signatures of Modified and Quantum Gravity'', organized by Christian Pfeifer
and Claus L\"ammerzahl.  LS was supported by the National SKA Program of China
(2020SKA0120300),  the National Natural Science Foundation of China (11975027,
11991053, 11721303), and the Max Planck Partner Group Program funded by the Max
Planck Society. 
\end{acknowledgement}

\bibliographystyle{abbrv}
\bibliography{refs}

\end{document}